\documentclass[10pt,letterpaper]{article}
\usepackage[top=0.85in,bottom=0.9in,left=1.25in,right=1in,footskip=0.75in]{geometry}

\usepackage{amsmath,amssymb}

\usepackage{changepage}

\usepackage{textcomp,marvosym}

\usepackage{cite}

\usepackage{nameref,hyperref}

\usepackage[nopatch=eqnum]{microtype}
\DisableLigatures[f]{encoding = *, family = * }

\usepackage[table]{xcolor}

\usepackage{array}
\usepackage{booktabs}

\newcolumntype{+}{!{\vrule width 2pt}}

\newlength\savedwidth

\raggedright
\usepackage[aboveskip=1pt,labelfont=bf,labelsep=period,justification=raggedright,singlelinecheck=off]{caption}

\makeatletter
\renewcommand{\@biblabel}[1]{\quad#1.}
\makeatother

\usepackage{lastpage,fancyhdr,graphicx}
\graphicspath{{./}}
\fancyheadoffset[L]{0.25in}
\fancyfootoffset[L]{0.25in}

\begin{document}
\vspace*{0.2in}

\begin{flushleft}
{\Large
\textbf{Large language model-assisted discovery of cohorts from scientific literature}
}
\newline
Moritz Sturm\textsuperscript{1,2\dag},
Lisa M. Berg\textsuperscript{1},
Inken Berg\textsuperscript{1},
Harishny Sarma\textsuperscript{1},
Jasmin Hartmann\textsuperscript{1},
Denissa Girschik\textsuperscript{1},
Gemma Roig\textsuperscript{2,3},
Christine M. Freitag\textsuperscript{1},
Andreas G. Chiocchetti\textsuperscript{1\dag}
\\
\bigskip
\textbf{1} Department of Child and Adolescent Psychiatry, Psychosomatic Medicine and Psychotherapy, University Hospital Frankfurt, Goethe University Frankfurt, Frankfurt am
Main, Germany
\textbf{2} Department of Computer Science, Goethe University Frankfurt, Frankfurt am
Main, Germany
\textbf{3} The Hessian Center for Artificial Intelligence (hessian.AI), Darmstadt, Germany
\\
\bigskip
$\dag$ Corresponding authors: \href{mailto:mosturm@uni-frankfurt.de}{mosturm@uni-frankfurt.de}; \href{mailto:andreas.chiocchetti@med.uni-frankfurt.de}{andreas.chiocchetti@med.uni-frankfurt.de}
\end{flushleft}

\section*{Abstract}
\setlength{\parindent}{0pt}
\textbf{Background}

Planning multi-study analyses requires identifying cohorts with the relevant participants, phenotypes, and data modalities. This process commonly relies on prior knowledge, cohort catalogues, and manual literature searches. We developed a complementary question-driven framework that searches relevant scientific literature and extracts explicit cohort names.

\textbf{Methods}

The framework first generates multiple PubMed queries from configurable vocabularies and templates and retrieves the resulting scientific literature automatically through the PubMed API. A large language model then screens the retrieved titles and abstracts and extracts explicit cohort names using a prompt tailored to the research question. The extracted names are deduplicated with human review. Configurable code, prompts, and example outputs are available at \url{https://gitlab.rz.uni-frankfurt.de/cap_molgenlab/literature-cohort-discovery}.

\textbf{Evaluation}

As a use case, we applied the framework to youth aggression genetics. From 5,400 generated PubMed queries, the framework retrieved 5,254 unique records and identified 188 candidate cohorts. Manual screening using predefined criteria, including participant age and genetic-data availability, retained 44 eligible cohorts. Automated LLM-based name extraction was within the agreement range of human annotators. We also searched four established cohort catalogues using the same research question. Their combined results contained 27 of the 44 eligible cohorts, while 17 were not returned by any cohort catalogue search.

\textbf{Conclusion}

The framework converts research-question-specific vocabulary into screenable cohort inventories via a large, automated literature search. It can be adapted across populations, phenotypes, data modalities, and study designs, and provides a literature-based complement to curated cohort catalogues.

\textbf{Keywords}

cohort discovery; dataset discovery; large language models; information retrieval; metadata; data catalogues

\section*{Introduction}
The reuse and joint analysis of existing human data have become increasingly important in population-based data science. A critical first step for researchers who conduct such analyses is to find suitable cohorts that contain individual-level data from the population of interest and include the relevant phenotypes.

In practice, information about human cohorts is typically fragmented across the internet, and researchers commonly rely on prior knowledge, exploratory web searches, or recommendations from colleagues to identify them \cite{zimmerman2007not, gregory2018eleven}. Although this expertise is valuable, it may be biased towards well-known cohorts, and the resulting set of cohorts can be difficult to audit or to reproduce for other researchers.

Prior work has therefore suggested approaches and frameworks for cohort selection analogous to systematic literature searches \cite{maelstromharmo,datagraphy}.
Such approaches nevertheless require suitable resources through which datasets can be discovered.

Several specialised cohort and data catalogues address this need. The Maelstrom Catalogue \cite{maelstrom_research_catalogue,bergeron2018fostering}, Atlas of Longitudinal Datasets \cite{atlas_longitudinal_datasets_2026}, Catalogue of Mental Health Measures \cite{CatMHM}, and dbGaP \cite{ncbi_dbgap,tryka2014ncbi} provide curated study-, measure-, and phenotype-level metadata. These catalogues are important discovery resources but differ in coverage and terminology. Moreover, they are generally organised around data resources and measures rather than individual research questions. Recent work in this field suggests that LLM-based systems can improve and complement dataset discovery within existing catalogues by matching the meaning of user queries to catalogue metadata \cite{green2026comparing}. However, these systems remain limited to resources represented in those catalogues.

Scientific publications provide a complementary source because they document cohorts in the context of their actual use.

Prior work has suggested dataset discovery based on citation contexts in publications \cite{scientificDatasetRecommendation, tan2025multi}, dataset discovery in topic-specific PubMed articles \cite{Tingay_Anastasiou_2018}, and, more recently, the use of large language models and retrieval-augmented generation to identify dataset mentions in scientific articles \cite{heddes,raging,ChatPD,datagath}. 

These studies demonstrate that scientific literature can support dataset extraction at scale, but to the best of our knowledge, none was developed to establish cohort discovery as a systematic framework for identifying suitable cohorts for a specific research question.

We therefore developed a configurable, LLM-assisted cohort-discovery pipeline that translates a research question into systematic PubMed searches and extracts explicitly named cohorts from abstracts and titles without requiring a predefined cohort list. 

We demonstrate the cohort discovery approach in genetic research on aggressive behaviour in children and adolescents. Genetic analyses of aggressive behaviour typically require samples substantially larger than most individual cohorts can provide \cite{Waltes2026,Baker2008,EAGLE2016,Tielbeek2017}, making cross-cohort research necessary. However, a unified resource of suitable cohorts for this research question is currently missing, making it a suitable use case for niche cohort discovery.

This use case also serves as a validation of the pipeline. The cohorts identified by the automated pipeline are evaluated through human screening against predefined eligibility criteria and compared with established cohort catalogues to examine overlap and complementary coverage.

Although the workflow operates on publications and cohort-level metadata, its target resources are human cohorts containing individual-level data. While we demonstrate and evaluate our approach through cohort discovery in youth aggression genetics, the PubMed search and LLM extraction are configurable for other populations, phenotypes, data modalities, and study designs.

\section*{Methods}

\subsection*{Study design and framework overview}
\begin{figure}[ht]
    \centering
    \includegraphics[width=0.5\textwidth]{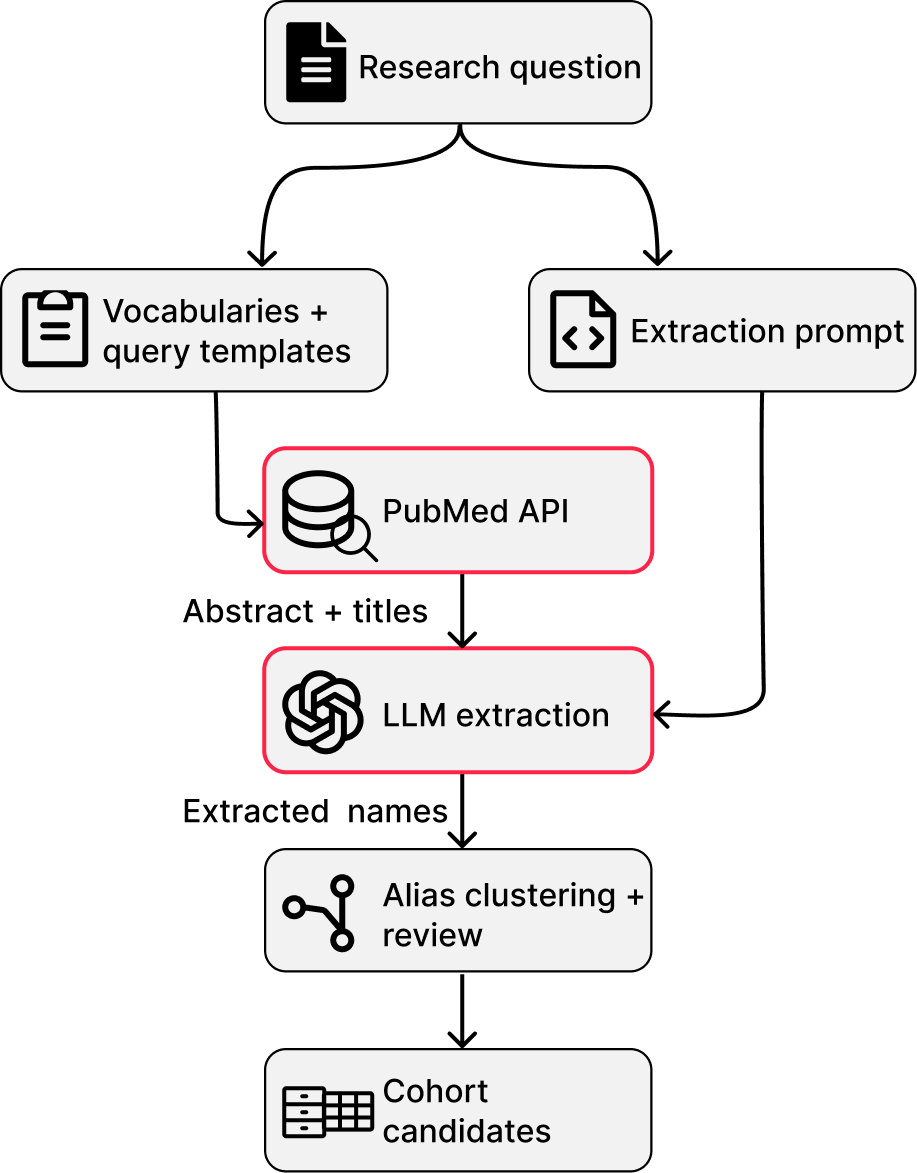}
    \caption{\textbf{Overview of the cohort-discovery workflow.} First, a research question is translated into question-specific vocabularies, PubMed query templates, and extraction instructions. Publications are then retrieved through the PubMed API, followed by large language model (LLM)-assisted extraction of cohort names. Finally, extracted name variants are clustered as potential aliases and manually reviewed to produce the candidate cohort inventory. Red outlines denote automated processing steps.}
    \label{fig:workflow}
\end{figure}

We developed a reusable workflow for identifying named cohorts relevant to a predefined research question. It comprises three stages: (i) retrieval of question-specific publications through the PubMed API \cite{PubMed}, (ii) LLM-assisted extraction of explicit cohort mentions \cite{GPT4o}, and (iii) deduplication of aliases and alternative names. The workflow takes configurable concept vocabularies, query templates, and extraction instructions as input (see Fig. \ref{fig:workflow}). It produces a list of
extracted, deduplicated cohort names, their observed name variants, and the PubMed records in which they were identified. Search terms and LLM instructions can be easily configured for application to other research questions and domains in the accompanying code \url{https://gitlab.rz.uni-frankfurt.de/cap_molgenlab/literature-cohort-discovery}.

\subsection*{Question-specific literature retrieval}
\label{sec:litrev}

The retrieval step automatically generates PubMed queries based on vocabularies relevant to the research question, such as the target population,
research domain, phenotype, data modality, and study design. These vocabularies are expanded across two configurable query templates:

\begin{verbatim}
population AND (domain OR phenotype) AND modality
population AND domain AND phenotype AND modality AND study-design
\end{verbatim}

The resulting queries are submitted via the PubMed API \cite{PubMed}.
The first template was chosen to be deliberately broad by requiring either the domain or
phenotype in order to reduce missed records caused by variable terminology. The
second template is designed to increase specificity of the search by requiring a more
restrictive match of domain, phenotype, and study design.
For each query, we retrieved the 50 highest-ranked titles and abstracts to balance coverage and computational feasibility. Records were deduplicated by PubMed ID before the LLM-assisted name extraction. 

\subsection*{LLM-assisted cohort-name extraction}
Each unique title and abstract was processed through the OpenAI API using the GPT-4o model snapshot \texttt{gpt-4o-2024-08-06} \cite{GPT4o}, a fixed prompt, and a structured JSON response schema. Only retrieved titles and abstracts were submitted, and therefore the workflow did not access participant-level data or full-text articles. The prompt first classified the record as \texttt{INCLUDE} or \texttt{DISCARD} according to the configured research direction and screening instructions. Included records were then screened for an explicit database, study, or cohort name. Only records labelled \texttt{INCLUDE} in which one or more explicit names were detected entered downstream processing.  

When a response contained multiple names, each was stored separately while retaining its source PMID. To reduce the effect of stochastic variation in LLM
outputs, the extraction step was repeated three times and the union of all extracted
PubMed ID--cohort name pairs across runs was used as input for the subsequent
deduplication step. An example of the extraction step and the prompt used for our youth aggression screening are shown in \nameref{sec:supp-example-extraction}. The LLM output was used to extract cohort names rather than to determine final eligibility. Those decisions remained subject to human verification.

\subsection*{Cohort alias and identity resolution}
To address naming inconsistencies and abbreviations, a semi-automatic deduplication step was introduced. Near-duplicate names were then clustered using Python-based string matching
(\texttt{difflib.get\_close\_matches}; similarity threshold = 0.8) \cite{PythonDifflib, gestaltdobb}. Each cluster was manually reviewed. Clusters identifying cohort names were assigned a canonical label, whereas those referring to other entities, such as diagnostic instruments, were removed.

We found human review to be critical for alias resolution because cohort names often appeared as abbreviations, alternative spellings, or nested substudies that could not be safely merged without contextual review. 

The final output of this step was an alias map linking canonical labels to their observed naming variants and to
the PubMed IDs in which these variants were identified. In our use case, for example, extracted variants such as \textit{ABCD study}, \textit{Adolescent Brain Cognitive Development study}, and \textit{Adolescent Brain Cognitive Development (ABCD) cohort} were resolved to the canonical label \textit{ABCD}.

\section*{Evaluation $\&$ Illustration}

\subsection*{Case-study search}

We evaluated the workflow by applying it to cohort discovery in youth aggression genetics. The search terms covered population (\textit{children, adolescents, youth, teenagers}); domain (\textit{mental health, conduct disorder, psychological development, behavioral problems, cognitive development}); phenotype (\textit{aggression, aggressive behavior, impulsive behavior, callous-unemotional traits, conduct disorder}); data modality (\textit{social, hormonal, endocrine, biobank, genotype, biological samples, phenotypic, imaging genetics, genetic}); and study design (\textit{longitudinal study, cohort study, case-control study, birth cohort, prospective study}). Applying the two query templates described in the \nameref{sec:litrev} subsection to these terms generated 5,400 PubMed queries. PubMed retrieval and automated screening began on 20 January 2025.

The PubMed API returned 261,621 query-level records, which were reduced to 5,254 unique records after PubMed ID deduplication. LLM-assisted extraction based on the prompt shown in Supplementary Fig.~\ref{fig:examplework} identified explicit cohort names in 476 records, and alias resolution yielded 188 candidate cohorts.

Each candidate was independently screened by two reviewers, resulting in 376 manual eligibility assessments. Cohorts were eligible if they (i) were currently accessible through documented data-access procedures; (ii) provided individual-level genetic data; (iii) included more than 1,000 participants; (iv) included participants younger than 18 years; and (v) assessed aggression-related phenotypes using validated questionnaires. Candidates were excluded if any criterion was not met. Reviewer disagreements were resolved by consensus, with adjudication by a third reviewer. Manual screening took place between 30 January and 25 July 2025 and required an estimated 400 person-hours.

Manual screening retained 44 eligible cohorts representing an aggregate reported sample of approximately 890,000 participants. This case study provided the setting for evaluating coverage relative to established cohort catalogues. Further methodological and catalogue details are reported in the accompanying data descriptor \cite{sturm2026global}. The catalogue of eligible cohorts is available on Zenodo \cite{sturm2026}.

\subsection*{Validation of automated name extraction}
\label{sec:llmextractionacc}
To assess LLM reliability, we compared its outputs with a manually annotated reference set of 199 randomly sampled records. Two human raters (R1, R2) independently labelled each record as either containing or not containing at least one relevant cohort. Both raters were presented with the same instructions provided to the LLM (Supplementary Fig.~\ref{fig:examplework}).

Because the task involves subjective judgements, including borderline resources and ambiguous acronyms, the two raters did not always agree. We therefore defined a conservative gold standard: an abstract was positive only when both raters labelled it as containing at least one relevant cohort, and records with disagreement (R1 $\neq$ R2) were excluded. Applying this definition, 176 of 199 abstracts (88.4\%) remained, consisting of 26 positives and 150 negatives.

Across all 199 abstracts, and treating R1 as the reference, R2 achieved an F1-score of 0.693 and a balanced accuracy of 0.877.
The LLM prediction for each record was defined by whether it extracted an explicit candidate name during the initial extraction step. These predictions were compared with the gold standard (agreement subset, n = 176) and with each rater's annotations separately.

We computed sensitivity (recall), precision, F1-score, and balanced accuracy from the numbers of true positives, false positives, false negatives, and true negatives. The metric definitions are provided in \nameref{sec:supp-extraction-validation}.

The results are summarised in Table~\ref{tab:llm_metrics}. Overall, the LLM operated within the agreement range of human annotators, which is consistent with recent work on LLMs in information-annotation tasks \cite{jensen2025,liu2025}. It aligned more closely with R1 (F1 = 0.821) than R1 aligned with R2 (F1 = 0.693), and its agreement with R2 (F1 = 0.648) was similar to human-human agreement. The model further achieved perfect precision (no false positives) but missed a subset of cohort mentions identified by both raters (recall = 0.769). 

Most false negatives occurred in studies at the boundary or outside our target domain, including adult samples, general well-being, or non-mental-health cohorts, where human raters tended to include any explicit study or cohort name. In contrast, the LLM followed a more conservative, topic-focused interpretation of the instructions. Individual examples are provided in \nameref{sec:supp-false-negatives}.

\begin{table}[!ht]
    \centering
    \caption{\textbf{Performance metrics for LLM-based database name extraction.} The generative pre-trained transformer (GPT) was compared against the gold standard (agreement between human rater 1 [R1] and human rater 2 [R2]) and against each individual rater. The last row shows the human-human baseline (R1 vs. R2).}
    \label{tab:llm_metrics}
    \resizebox{\textwidth}{!}{%
    \begin{tabular}{lrrrrrr}
        \hline
        Comparison & n & Positives & Precision & Recall & F1-score & Balanced accuracy \\
        \hline
        GPT vs. GOLD (R1 and R2 agree) & 176 & 26 & 1.000 & 0.769 & 0.870 & 0.885 \\
        GPT vs. R1                     & 199 & 30 & 0.885 & 0.767 & 0.821 & 0.874 \\
        GPT vs. R2                     & 199 & 45 & 0.885 & 0.511 & 0.648 & 0.746 \\
        R1 vs. R2 (human-human)        & 199 & 30 & 0.578 & 0.867 & 0.693 & 0.877 \\
        \hline
    \end{tabular}
    }
\end{table}

To assess the reproducibility of name extraction, we compared two independent runs across all 5,254 unique title--abstract pairs. Each run produced a set of PMIDs for records in which the pipeline detected explicit names. Agreement between runs was quantified using the Jaccard similarity coefficient,
\[
    J(A, B) = \frac{|A \cap B|}{|A \cup B|},
\]
where A and B are the PubMed ID sets from each run. The resulting coefficient, J(A, B) = 0.85, indicated that the two runs generally identified the same records as containing explicit names.

\subsection*{Comparison with established cohort catalogues}
Accordingly, this analysis evaluates overlap and complementary coverage.

To assess whether established catalogues could reproduce the cohorts found in our use case, we compared the literature-derived cohort inventory (188 candidates; 44 eligible) with documented searches of the Atlas of Longitudinal Datasets \cite{atlas_longitudinal_datasets_2026}, Catalogue of Mental Health Measures (CatMH) \cite{CatMHM}, Maelstrom Catalogue \cite{maelstrom_research_catalogue,bergeron2018fostering}, and dbGaP \cite{ncbi_dbgap,tryka2014ncbi}. Because the platforms differed in scope and search functionality, the case-study queries and inclusion criteria were translated into catalogue-specific search strategies. The catalogue searches were conducted on 23 July 2026.

The complete catalogue-specific search strategies are provided in \nameref{sec:supp-catalogue-searches}.

\begin{figure}[!ht]
    \centering
    \includegraphics[width=0.723\textwidth]{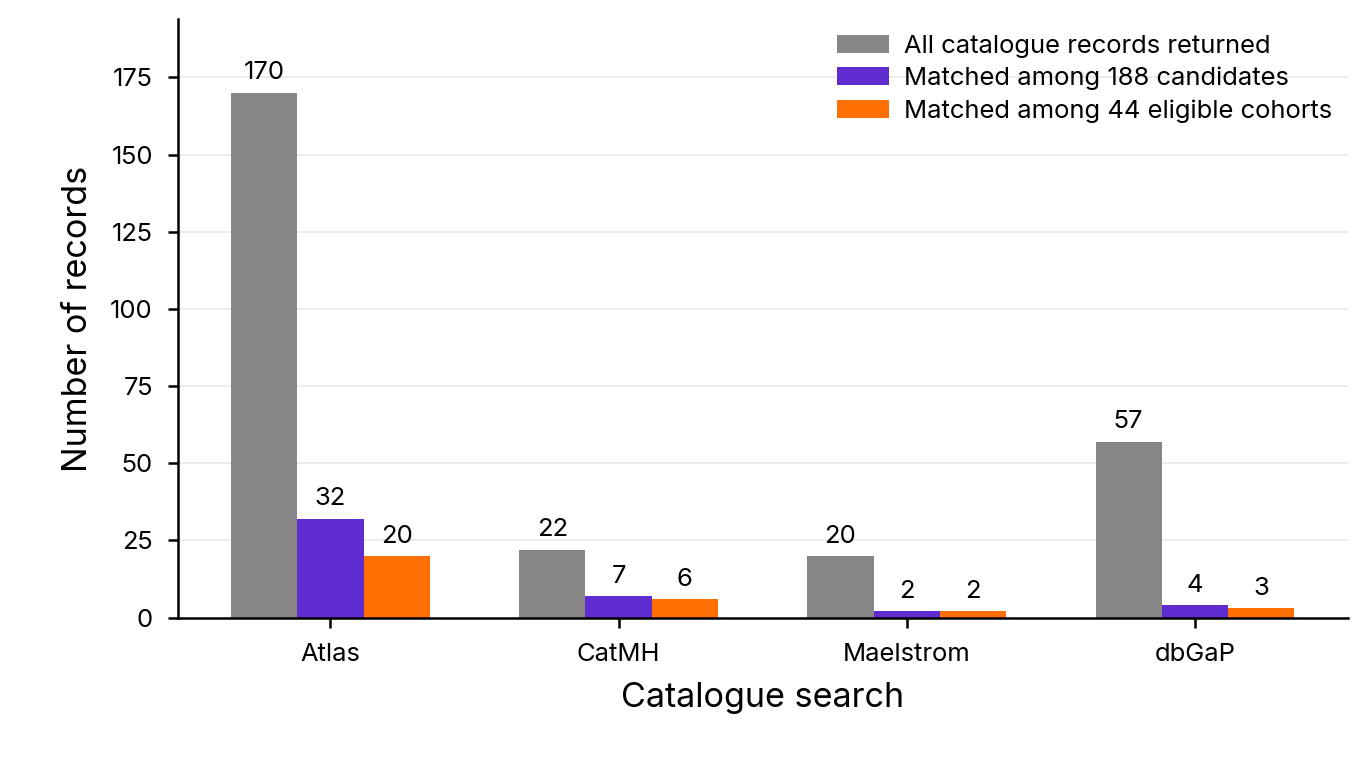}
    \caption{\textbf{Catalogue search returns and overlap with the
    literature-derived cohort inventory.}
    Grey bars show the cohort identities retrieved from each catalogue;
    blue and orange bars show matches to the 188 literature-derived candidates
    and 44 eligible cohorts, respectively. Counts are catalogue-specific, and
    the same cohort may occur in more than one catalogue. CatMH denotes
    Catalogue of Mental Health Measures; dbGaP, Database of Genotypes and
    Phenotypes.}
    \label{fig:catalogue_validation}
\end{figure}

These searches returned 170 Atlas, 22 CatMH, 20 Maelstrom, and 57 unique dbGaP identities (Fig.~\ref{fig:catalogue_validation}). Records were linked to the literature-derived inventory using abbreviations and reviewed name variants. After deduplication across catalogues, their union contained 253 distinct cohort identities. Catalogue-only cohorts were not screened against the case-study criteria.
The catalogue union contained 40 of the 188 candidates (21.3\%), including 27 of 44 eligible cohorts (61.4\%) and 13 of 144 excluded candidates (9.0\%). 

The documented catalogue result sets showed little overlap, with 237 of 253 unique cohorts (93.7\%) occurring in only one catalogue. No single catalogue reproduced the
question-derived eligible set, while 17 eligible cohorts identified through the
literature workflow were not returned by any of the four documented catalogue
searches. These findings support literature mining as a complementary cohort
discovery route alongside established catalogues.

\section*{Discussion}

We present a configurable, LLM-assisted workflow for discovering named population cohorts from scientific literature.

Previous work has already shown that cohort and dataset mentions can be extracted from scientific publications \cite{Tingay_Anastasiou_2018,heddes,raging,ChatPD,datagath}. We extend this principle from mention extraction to question-driven cohort discovery. The main methodological advance of the proposed work therefore lies in integrating automated literature retrieval and LLM-based name extraction into a configurable workflow.

We chose the niche problem of cohort discovery in youth aggression genetics to illustrate and validate our approach. 
In this case study, LLM-assisted extraction
reduced 5,254 unique publication records to 188 candidates for cohort screening without
requiring a predefined list of cohort names. At approximately USD 10, LLM
processing was very cost-effective compared with the substantial manual labour that
would have been required otherwise.

The validation results further supported the utility of automated LLM extraction.
Automated extraction performed within the range of human inter-rater agreement
observed in our use case, consistent with previous work \cite{jensen2025,liu2025}. The imperfect
agreement between human raters shows that identifying cohort names in 
titles and abstracts involves genuine interpretive uncertainty. In addition, we showed that the LLM consistently identified cohort mentions in the same titles and abstracts across repeated runs.

We implemented and evaluated GPT-4o, the newest available LLM at the time of
screening. However, the methodological contribution is the configurable workflow rather
than the use of a specific LLM, and the model can be updated to more modern versions in the accompanying code. Although this was not explicitly tested, we expect
further advances in LLMs to benefit rather than impair the automated extraction step.
Future implementations could extend the workflow by additionally using an LLM to aggregate cohort-specific online information for eligibility assessment. We deliberately retained fully human eligibility assessment in this study to provide an independent validation setting for the proposed pipeline.

The comparison with established cohort catalogues showed that the value of the proposed workflow lies not in replacing curated catalogues, but in addressing a different discovery problem. Catalogue searches identify resources through structured metadata and platform-specific indexing, whereas the literature-based workflow identifies cohorts through publications in which they have been used for a question-relevant purpose. The documented searches demonstrated substantial fragmentation across catalogue resources, and no individual catalogue search reproduced the same list of cohorts as the literature-driven approach. At the same time, 17 eligible cohorts identified through the literature workflow were not returned by any of the four documented searches.

The two approaches can therefore be considered complementary, and we argue that our proposed workflow is best used to broaden the candidate space of cohorts for a specific research question. Our workflow may also be used to support the development and completion of cohort
catalogues by identifying cohorts for possible addition to them.

The findings should be interpreted in light of the study's scope. The workflow was evaluated in one application, and performance with other populations, phenotypes, or terminology remains to be established. Retrieval was restricted to PubMed because its API supported automated execution of the generated queries. This provides a reproducible source for biomedical literature, although cohorts documented mainly in other disciplines or indexing systems may be less visible. Identification also depends on explicit cohort naming and information present in titles and abstracts.

Finally, catalogue platforms required different search strategies because their interfaces and metadata structures differ in practice. Catalogue-only identities were not screened against all case-study criteria because such rescreening was not feasible. The comparison therefore characterises overlap and complementarity among the documented searches. 
Another limitation is the temporal gap in the evaluations. PubMed retrieval and manual screening were conducted in 2025, whereas the catalogue searches were conducted in July 2026. Changes in catalogue contents or cohort documentation during this interval may have affected the observed overlap.
These limitations bound the reported evaluation but do not alter the central finding
that question-driven, LLM-assisted literature extraction adds a practical discovery route
alongside established cohort catalogues.

\section*{Conclusions}
We developed a configurable, LLM-assisted and human-reviewed workflow that addresses a central challenge in population-data research: identifying cohorts relevant to a scientific question without relying on a predefined cohort list. By combining systematic PubMed retrieval, contextual cohort-name extraction, and canonical identity resolution, the workflow converts a large publication corpus into a screenable cohort inventory. Its application to youth aggression genetics demonstrated practical feasibility and produced an open catalogue of 44 eligible cohorts. Comparison with four established catalogues showed that literature- and catalogue-based searches provide complementary routes to cohort discovery. The workflow can also be used to identify candidates for cohort catalogue curation. Further applications should establish performance across other populations, phenotypes, data modalities, and research domains.

\section*{Data availability statement}
\label{sec:Dataavail}
The youth-aggression cohort catalogue and accompanying documentation are publicly available on Zenodo at \url{https://doi.org/10.5281/zenodo.18216492}.

\section*{Data provenance statement}
\label{sec:code}
The publication records originated from PubMed and were retrieved through the PubMed API using the generated case-study queries beginning on 20 January 2025. Records were deduplicated by PMID before their titles and abstracts were processed in three independent LLM runs. Extracted cohort names were combined across runs, clustered using string similarity, and manually reviewed to resolve aliases and assign canonical labels. Candidate cohorts were then independently assessed by two reviewers, with disagreements resolved by consensus and adjudication by a third reviewer. The external comparison used documented searches of the Atlas of Longitudinal Datasets, Catalogue of Mental Health Measures, Maelstrom Catalogue, and dbGaP conducted on 23 July 2026. Catalogue identities were linked using abbreviations and reviewed name variants and then deduplicated across catalogues. Source PMIDs, processing outputs, and the configurable implementation are available at \url{https://gitlab.rz.uni-frankfurt.de/cap_molgenlab/literature-cohort-discovery}; the resulting cohort catalogue is archived on Zenodo at \url{https://doi.org/10.5281/zenodo.18216492}.

\section*{Ethics statement}
Ethical approval was not required because this methodological study analysed publicly available publication records and cohort-level metadata and did not involve research participants, participant-level data, or identifiable personal information.

\section*{AI disclosure statement}
GPT-4o (model snapshot \texttt{gpt-4o-2024-08-06}) was used through the OpenAI API as a research instrument to screen PubMed titles and abstracts and extract explicit cohort names. Only publication titles and abstracts were submitted to the model, and no participant-level data or personal identifiers were processed. Model outputs generated candidates rather than final eligibility decisions. Because the extraction model operated on publication-level metadata rather than participant-level predictions, a demographic fairness audit was not applicable. Potential biases related to PubMed coverage, publication practices, and explicit cohort naming are addressed in the Discussion. During preparation of the work, OpenAI Codex and GPT-5 were used to assist with manuscript editing and code restructuring and development. The authors designed the study and methodology, made all screening and analytical decisions, reviewed and verified all AI-assisted outputs, and take full responsibility for the code, analyses, and manuscript.

\section*{Author contributions}
M.S. designed the methodology, implemented the analysis code, evaluated the technical validity of the pipeline, and coordinated the screening process. M.S. and L.M.B. drafted the manuscript. M.S., L.M.B., I.B., H.S., J.H., and D.G. conducted the manual screening of cohort resources and contributed to metadata extraction. I.B. contributed to the initial conceptualisation of the project. G.R. advised on the design of the model evaluation and provided feedback on the manuscript. C.M.F. contributed clinical and methodological expertise, refined the clinical inclusion criteria for the search, and provided critical feedback on manuscript drafts. A.G.C. proposed the research idea, guided the research, contributed to the evaluation of the automated pipeline, and provided feedback on the manuscript. C.M.F. and A.G.C. jointly secured funding for the project. All authors reviewed the final manuscript and approved the submitted version.

\section*{Acknowledgements}
We thank Dino Milanovic for guidance on the design of the methodological workflow figure. This work was funded by the Deutsche Forschungsgemeinschaft (DFG, German Research Foundation) -- Project-ID 512007073 -- TRR 379.

\section*{Conflicts of interest}
The authors declare no conflicts of interest related to the submitted work.

\section*{Abbreviations}
AI, artificial intelligence; API, application programming interface; CatMH, Catalogue of Mental Health Measures; dbGaP, Database of Genotypes and Phenotypes; DFG, Deutsche Forschungsgemeinschaft; GPT, generative pre-trained transformer; JSON, JavaScript Object Notation; LLM, large language model; PMID, PubMed identifier.

\section*{Supplementary Appendix}
Supplementary Appendix 1. Structured extraction response, example extraction workflow, extraction-validation details, and documented catalogue search strategies.

\bibliography{refs_harmonized}

\clearpage

\section*{Supplementary Appendix 1}

\textbf{Contents}
\begin{itemize}
    \item Supplementary Section S1. Structured extraction response
    \item Supplementary Section S2. Example extraction workflow
    \item Supplementary Section S3. Extraction-validation details
    \item Supplementary Section S4. Documented catalogue search strategies
\end{itemize}

\setcounter{figure}{0}
\renewcommand{\thefigure}{S\arabic{figure}}

\subsection*{Supplementary Section S1. Structured extraction response}

The fixed response schema used three fields:

\begin{verbatim}
{
  "decision": "INCLUDE" or "DISCARD",
  "category": 0, 1, or 2,
  "details": "free-text explanation or extracted cohort name(s)"
}
\end{verbatim}

Category 0 denoted no named database, study, or cohort; category 1 denoted one
or more explicit names; and category 2 denoted a referenced but unnamed
resource. Only responses with \texttt{decision = INCLUDE} and
\texttt{category = 1} were retained. The complete fixed prompt is archived in
the accompanying code repository.

\subsection*{Supplementary Section S2. Example extraction workflow}
\label{sec:supp-example-extraction}

\begin{figure}[ht]
    \centering
    \includegraphics[width=0.95\textwidth]{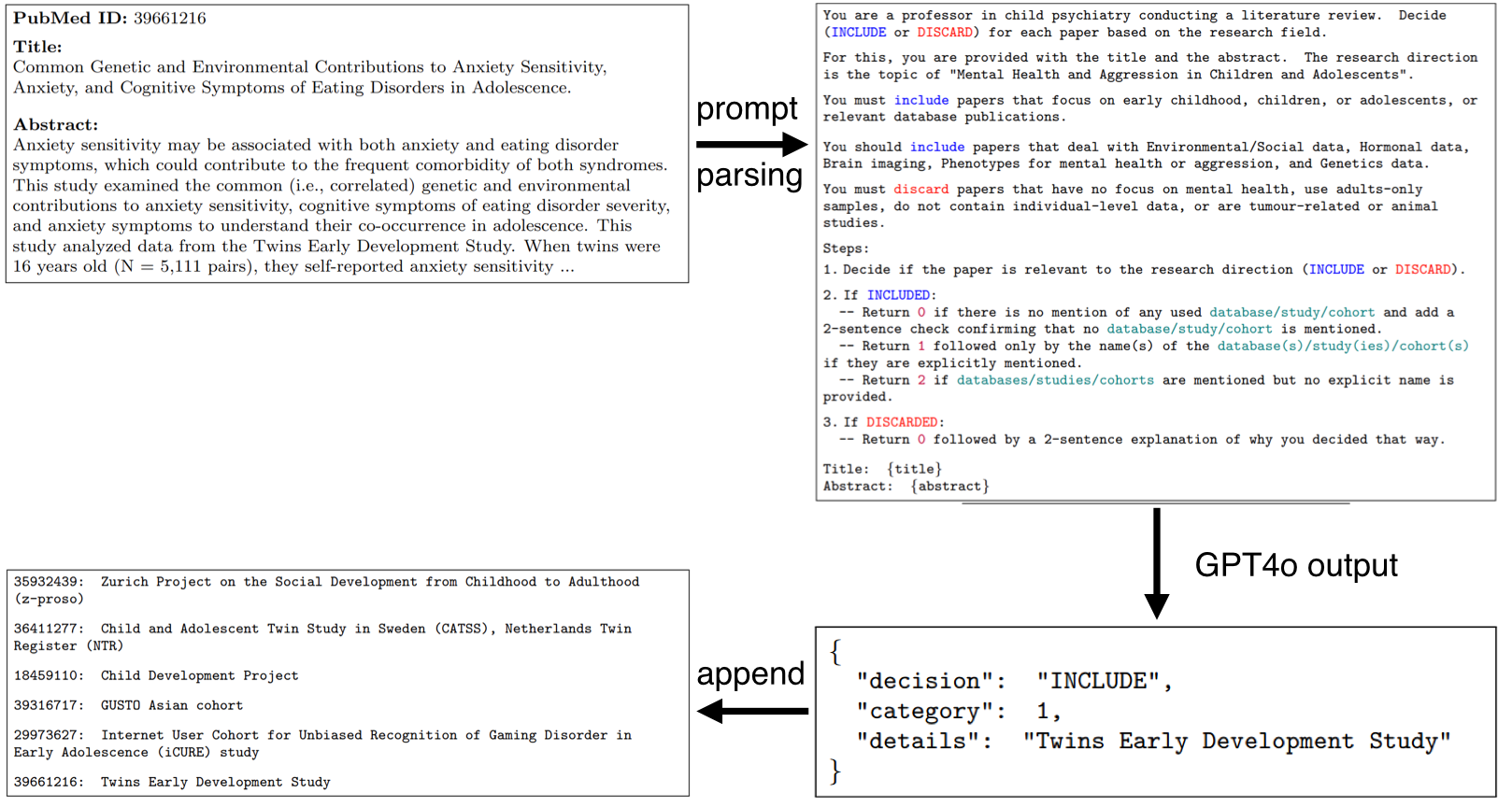}
    \caption{\textbf{Example workflow for AI-assisted cohort-name extraction
    in the youth-aggression application.} Titles and abstracts retrieved
    through the PubMed API were inserted into a structured prompt and submitted
    to the GPT-4o API. The model returned a structured JSON response containing
    an \texttt{INCLUDE}/\texttt{DISCARD} decision and any extracted cohort
    names. Extracted names were linked to the corresponding PMID and added to
    the candidate list. The prompt, publication record, and model output shown
    are based on real instances; the abstract for PMID 39661216 was shortened
    for display. The complete prompt and implementation are available in the
    accompanying code repository (see \nameref{sec:code}).}
    \label{fig:examplework}
\end{figure}

\newpage

\subsection*{Supplementary Section S3. Extraction-validation details}
\label{sec:supp-extraction-validation}

\subsubsection*{Classification metrics}

Classification performance was calculated from true positives (TP), false
positives (FP), false negatives (FN), and true negatives (TN):

\begin{align*}
\mathrm{Recall} &= \frac{\mathrm{TP}}{\mathrm{TP} + \mathrm{FN}}, &
\mathrm{Precision} &= \frac{\mathrm{TP}}{\mathrm{TP} + \mathrm{FP}}, \\
F_1 &= 2 \cdot
\frac{\mathrm{Precision} \cdot \mathrm{Recall}}
{\mathrm{Precision} + \mathrm{Recall}}, &
\mathrm{Balanced\ Accuracy} &= \frac{1}{2} \left(
        \frac{\mathrm{TP}}{\mathrm{TP} + \mathrm{FN}} +
        \frac{\mathrm{TN}}{\mathrm{TN} + \mathrm{FP}}
    \right).
\end{align*}

\subsubsection*{Examples of false-negative records}
\label{sec:supp-false-negatives}

Most false negatives occurred in studies at the boundary or outside the target
domain. Examples included adult follow-up cohorts originally recruited in
youth (VINGO, PMID 39595714; LONGSCAN, PMID 35156437), cohorts mentioned in a
record about happiness across the lifespan (ABCD, Add Health, and UK Biobank;
PMID 37828061), and a non-psychiatric genetic biobank for Marfan syndrome
(PMID 22330841).

\subsection*{Supplementary Section S4. Documented catalogue search strategies}
\label{sec:supp-catalogue-searches}

\subsubsection*{Atlas of Longitudinal Datasets}

Two full Boolean expansions were used:

\begin{verbatim}
(children OR adolescents OR youth OR teenagers) AND
("mental health" OR "conduct disorder" OR "psychological development" OR
 "behavioral problems" OR "cognitive development" OR aggression OR
 "aggressive behavior" OR "impulsive behavior" OR
 "callous-unemotional traits") AND
(social OR hormonal OR endocrine OR biobank OR genotype OR
 "biological samples" OR phenotypic OR "imaging genetics" OR genetic)

(children OR adolescents OR youth OR teenagers) AND
("mental health" OR "conduct disorder" OR "psychological development" OR
 "behavioral problems" OR "cognitive development") AND
(aggression OR "aggressive behavior" OR "impulsive behavior" OR
 "callous-unemotional traits" OR "conduct disorder") AND
(social OR hormonal OR endocrine OR biobank OR genotype OR
 "biological samples" OR phenotypic OR "imaging genetics" OR genetic) AND
("longitudinal study" OR "cohort study" OR "case-control study" OR
 "birth cohort" OR "prospective study")
\end{verbatim}

The documented filters selected accessible or contact-based access,
mental-health or psychological-measure data, biobank/cohort/registry designs,
ages 0--18 years, and sample sizes above 1,000.

\subsubsection*{Catalogue of Mental Health Measures}

The documented facets selected biomarkers, recruitment ages from birth to 18
years, and sample size of at least 1,000.

\subsubsection*{Maelstrom Catalogue}

The documented filters required maximum age below 18 years, more than 1,000
participants, and possible data access.

\subsubsection*{dbGaP}

\begin{verbatim}
(child OR adolescent OR youth) AND
(aggression OR "aggressive behavior" OR "conduct disorder" OR
 "behavioral problems" OR "impulsive behavior" OR
 "callous-unemotional traits")

(child OR adolescent OR youth) AND
("mental health" OR "conduct disorder" OR "behavioral problems") AND
(aggression OR "aggressive behavior" OR "callous-unemotional traits")
\end{verbatim}

\end{document}